# Effects of Applied Load on Formation and Reorientation of Zirconium Hydrides: A Multiphase Field Modeling Study

Alireza Toghraee[a], Jacob Bair[b,c], and Mohsen Asle Zaeem[a]*

[a]*Department of Mechanical Engineering, Colorado School of Mines, 1500 Illinois St., Golden, Colorado, USA*

[b]*Department of Mechanical Engineering, Oklahoma State University, 201 GAB, Stillwater, Oklahoma, USA*

[c]*Pacific Northwest National Laboratory, 906 Battelle Boulevard, Richland, Washington, USA*

## Abstract

The effects of applied load on the formation and reorientation of $\delta$ hydrides in $\alpha$-zirconium matrix are studied by a multiphase field model. Simulations with only one hydride seed indicate that the hydride reorientation occurs only when a tensile or compressive strain of ~0.02 is applied along $[11\bar{2}0]$ (hoop or circumferential direction) or $[0001]$ (radial direction aligned with the basal pole direction), respectively. However, two-seed and multi-seed hydride simulations show the significant influence of neighboring hydrides on the required load for reorientation, reducing it by more than one order of magnitude and making it comparable to the experimental results. Through phase-field simulations, this work suggests that the hydride reorientation happens more easily when clusters of hydrides are present, and the required external load for reorientation is within the elastic limit of the cladding material.

* Corresponding author; E-mail address: zaeem@mines.edu (M. Asle Zaeem).

## 1. Introduction

Zirconium (Zr) nuclear fuel claddings are constantly being water-cooled during operation and in the storage period. Reaction of water and claddings causes oxidation and releases hydrogen (H) atoms, some of which enter the cladding and cause formation of Zr hydrides. During the operation, nuclear fuel claddings are under constant hoop stresses caused by a combination of the pressure created by fission gases within the claddings and the helium fill gases in the cladding tubes. Other stresses occur during the transfer of spent nuclear fuel rods from operation to the storage, where the temperature changes and the internal pressure may cause re-precipitation and reorientation of Zr hydrides. The reorientation of hydrides has a significant effect on the mechanical behavior of the claddings and may cause facture and failure of the claddings, and this has been the subject of various experimental studies [1-5].

Claddings are fabricated to create a texture that promotes mostly circumferential orientations of hydride platelets [6]; however, reorientation of hydrides to radial direction has often caused a brittle fracture through the thickness of the cladding by a process known as Delayed Hydride Cracking (DHC). DHC is known as the most limiting factor to the lifetime of nuclear fuel rod claddings [7, 8]. Previous experimental studies indicate that understanding the effects of external loads on the formation and reorientation of $\delta$ hydrides in cladding materials is important in operating conditions, during transferring of the used fuel rods and also during the long-term storage [8].

The formation and shape evolution of stable $\delta$ hydrides in Zr alloys also depend on the metastable $\zeta$ and $\gamma$ phases [9, 10]. In a previous work, we developed a multiphase field model and studied the effects of metastable $\zeta$ and $\gamma$ phase hydrides in Zr-based nuclear fuel rod claddings without considering the effects of external loadings [11]. Although some phase-field simulations have been used previously to study the metastable $\zeta$ and $\gamma$ phases under various applied loads [12-14], there are only few works studied the effects of external loads (strains) on the stable $\delta$ phase [15]. With the exception of our recent work, there is only one phase-field modeling study that included the $\delta$ phase and focused on the effects of interface energies and temperature gradients in which the metastable phases were ignored and considered a direct formation path from $\alpha$-Zr matrix to $\delta$ phase hydrides [16].

Significant differences in elastic properties of different hydride phases and their stress-free strains indicate that applied loads will affect the formation and shape evolution of these phases. The metastable phases were shown to affect the formation and shape evolution of $\delta$ hydrides without an applied load [11], therefore it is reasonable to assume that they also can play a role in presence of external loads. In addition, the external loads can also affect the shape and orientation of the $\delta$ hydrides after they are formed, which this usually happens during the storage period and triggers the reorientation of the $\delta$ hydrides. Significant differences in the evolution of different hydride phases under various applied loads could lead to the discovery of preferable pre-stress treatments or improved operating conditions to avoid or mitigate the reorientation of hydrides. In this work, we first study the formation and reorientation of multiphase hydrides (metastable phases to stable $\delta$ phase) under applied load, and then we investigate the reorientation of the already formed stable $\delta$ hydrides under applied load. Furthermore, the effect of configuration and distance of hydrides from each other on the required load for reorientation is investigated.

## 2. Multiphase Field Model and Simulation Details

In the current study, the effect of applied load on evolution and reorientation of Zr-hydrides is investigated. Fig. 1 shows the schematic orientation of hydrides with respect to the Zr cladding. In order to simulate the cross section of a cladding, a two-dimensional (2D) domain is considered where the x-axis is in $\left[11\bar{2}0\right]$ direction and y-axis is in $\left[0001\right]$ direction (normal to basal plane in HCP Zr). $\left[0001\right]$ direction is typically oriented close to the radial direction in a cladding tube thus the y-axis will be noted as the radial direction, and the x-axis will be noted as the hoop or circumferential direction.

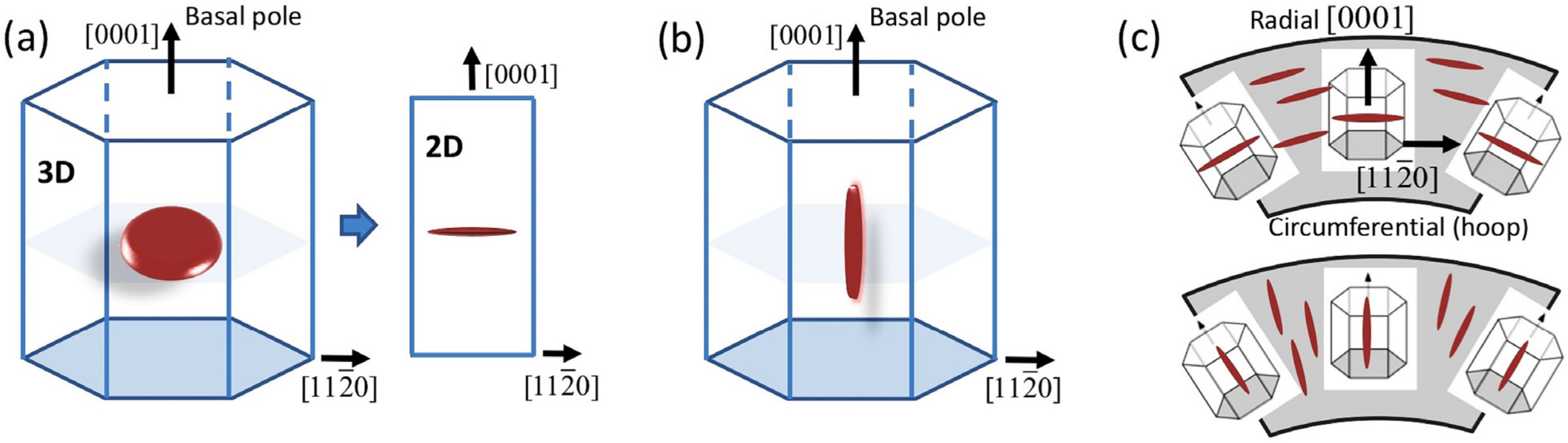

Fig. 1. Schematic view of $\delta$-hydride inside Zr cladding: a) 3D and 2D views before reorientation, b) 3D view after reorientation, and c) a 2D cut of the cladding shell showing the hydrides before (top) and after (bottom) reorientation.

Following our recent work [11], the multiphase field model uses one conserved phase-field variable to control the hydrogen concentration $C$, and six non-conserved structural field variables $\eta_i$, one for $\alpha$-Zr, one for $\zeta$-$Zr_2H$, one for $\delta$-$ZrH_{1.5+x}$, and three for the three eigenstrain variants of $\gamma$-ZrH. The total free energy of the system is defined as the summation of the chemical free energy and the elastic free energy:

$$F = F_C + F_{el}, \tag{1}$$

where $F_C$ is the total chemical free energy and $F_{el}$ is the elastic free energy:

$$F_C = \int_V \left[ f(C,\eta_i,T) + \sum_{i,j}^{n} \frac{K_{ij}}{2} \left|\nabla \eta_i\right| \left|\nabla \eta_j\right| \right] d\vec{r} , \tag{2}$$

$$F_{el} = \int_V f_{el}\, d\vec{r} = \frac{1}{2}\int_V \sigma_{ij}\varepsilon_{ij}^{el}\, d\vec{r} = \frac{1}{2}\int_V C_{ijkl}^{tot}\varepsilon_{kl}^{el}\varepsilon_{ij}^{el}\, d\vec{r} , \tag{3}$$

$$\varepsilon_{ij}^{el}(\vec{r}) = \varepsilon_{ij}^{tot}(\vec{r}) - \varepsilon_{ij}^{00}(\vec{r}) = \varepsilon_{ij}^{tot}(\vec{r}) - \sum_{n=1}^{6}\varepsilon_{ij}^{00}(n)\eta_n^2(\vec{r}) = \frac{1}{2}\left( \frac{\partial u_i(\vec{r})}{\partial r_j} + \frac{\partial u_j(\vec{r})}{\partial r_i} \right) - \sum_{n=1}^{6}\varepsilon_{ij}^{00}(n)\eta_n^2(\vec{r}) , \tag{4}$$

$$C_{ijkl}^{tot} = \sum_{n=1}^{6} \eta_n C_{ijkl}^{n} . \tag{5}$$

In the mentioned equations, $f$ is the chemical free energy density of the bulk, $K_{ij}$ is related to the interfacial free energy between the matrix and precipitates, $T$ is the temperature, $\sigma_{ij}$ is the stress, $C_{ijkl}$ is the elastic tensor, $\varepsilon_{ij}^{el}$ is the elastic strain, and $\varepsilon_{ij}^{00}$ is the stress free transformation strain for each orientation variable. The value of $n$ varies between 1 and 6 summing the values for all phases ($\alpha,\zeta,\gamma_1,\gamma_2,\gamma_3,$ and $\delta$). The displacements are represented by $u_i$. $\zeta$ phase is coherent with the $\alpha$-Zr matrix and has the same crystallographic orientation. $\gamma$ and $\delta$ phases have a $(111)_{\gamma,\delta} \| (0001)_\alpha$ plane relationship and a $[1\bar{1}0]_{\gamma,\delta} \| [11\bar{2}0]_\alpha$ directional relationship with the $\alpha$-Zr matrix. These crystallographic relationships lead to stress-free strains with no shear components. The derivation of these strains is given in details in a work by Carpenter for $\gamma$ and $\delta$ phases [17], and by Thuinet et al. for the $\zeta$ phase [13, 18]. The elasticity tensor transitions smoothly between phases using the non-conserved order parameters, by multiplying the elasticity tensor of each phase by its associated order parameter, as shown in Eq. (5). In this model, all interfaces are controlled through the non-conserved order parameters. Gibbs free energies from Christensen et al. are used [19] to present the formation of different phases.

Lattice mismatch strains caused by hydrogen diffusion through the matrix are not included in this model, which may affect the diffusion in the model but should not considerably affect the elastic responses of each phase. No defect (cracks, etc.) is considered in this work so high stress concentrations are not expected anywhere in the domain of study, therefore ignoring the lattice mismatch strains will not have any noteworthy effect on the results. Other recent two-phase models have also neglected these strains [10, 13] for similar reasons. In larger simulation domains, when diffusion could become a significant limiting factor on hydride growth, or in simulations with stress concentrators such as cracks, the model needs to include the lattice mismatch strains [12].

The maximum hydrogen solubility, used in calculating the chemical bulk free energy of the $\alpha$ phase as described in [11], is described using the equation given by Une et al. in 2009 defining the terminal solid solubility of precipitation of hydrides in wtppm [20]:

$$C_\alpha = 3.27\times 10^4 \exp\left(\frac{-25,042}{RT}\right). \tag{7}$$

The governing equations of the multiphase field model include evolution equations for the non-conserved field variables, Fick's diffusion equation for the conserved diffusion of hydrogen, and the mechanical equilibrium equation to control the evolution of stresses and strains:

$$\frac{\partial \eta_i}{\partial t} = -L_i\left(\frac{\partial f}{\partial \eta_i} + \frac{\partial f_{el}}{\partial \eta_i} - \sum_{i,j}^{n} K_{ij}\nabla^2 \eta_i\right), \tag{8}$$

$$\frac{\partial C}{\partial t} = M\nabla^2\left(\frac{\partial f}{\partial C}\right), \tag{9}$$

$$\nabla\sigma_{ij} = 0 = \nabla(C_{ijkl}\varepsilon_{kl}), \tag{10}$$

where $L_i$ is the structural relaxation coefficient, and $M$ is the mobility.

The material properties and model parameters used in this work are given in Table 1, where the other variants of stress-free strain of $\gamma$ hydride are obtained by rotating the given stress-free strains about the z axis (the [0001] direction in the $\alpha$-Zr matrix). The usual average grain size of cladding is in the micron length scale (e.g., 5 μm [21]), and our domain of study is sub-micron. Also, the manufacturing process of claddings creates a texture that promotes similar orientations for $\alpha$-Zr grains which makes the basal pole of HCP crystal aligned with the radial direction. Thus, single crystal simulations should be sufficient to provide necessary knowledge on the process and mechanisms of hydride reorientation in Zr claddings. Constants $A_\alpha$, $A_\zeta$, $A_\gamma$, and $A_\delta$ which are used to set the curvatures of the parabolic chemical free energy densities and the process used to determine the gradient energy terms and the barrier terms are explained in details in our recent work [11].

The governing equations, Equation (8), (9) and (10), are solved using the weak forms through the finite element implementations in the Multiphysics Object Oriented Simulation Environment (MOOSE) developed by Idaho National Laboratory [22, 23]. Neumann boundary conditions were used for all phase-field variables and Dirichlet boundary conditions were applied on displacement variables to create the desired applied strains. The initial hydrogen concentration in the matrix was set at 0.03 in the multiphase simulations and 0.01 for the case of stable $\delta$ hydrides accounting for consumption of hydrogen to form $\delta$ hydrides. The domain of study is a section of cladding bounded in radial and circumferential direction and along the length of the cladding (Fig. 1). This makes the thickness very large in comparison with other dimensions of the domain, and the strain in the direction normal to the plane can be neglected, thus, plane strain conditions were considered for our 2D simulations. An adaptive mesh algorithm was used to reduce the computational usage, and after a mesh convergence study, the smallest mesh size of 0.1 nm at the interfaces was chosen, and the mesh size elsewhere could increase to a size of 1.6 nm. Small mesh sizes were necessary due to the small interface energies between several of phases. These small interface energies led to small interface widths, which in turn required a very small mesh. The algorithms controlling the mesh adaptivity as well as the adaptive time stepping are described in detail elsewhere [11]. Simulations were run using the actual material properties to obtain quantitative results.

## 3. Results and Discussion

First, we consider the condition that a hydride starts to form and evolve from its metastable phase without and with an applied load. A single seed of metastable $\zeta$ phase was placed in the middle of the 2D domain at the $\left(10\bar{1}0\right)$ plane; $\left(10\bar{1}0\right)$ plane is often seen in the cross section of cladding tubes. Random nucleation of the other hydride phases was also considered. Results in Fig. 2a show the evolution of the single $\zeta$ seed into the $\delta$ phase under no applied load. The nucleation for $\gamma$ and $\delta$ hydrides is controlled using the probability of formation of a nucleus by the classical nucleation theory [11]. This sets the center of the nucleus, and in some cases a nucleus can be formed entirely inside the parent phase, or in other cases a nucleus may be formed at the interface between phases. The simulation results in Fig. 2a show that with no applied load no reorientation occurs.

**Table 1.** Material properties and model parameters used in simulations for all phases. Elastic constants, Bulk ($B$) to shear modulus ($G$) ratio from DFT and experiment [24-27], Interfacial energies from Thuinet or approximated [28], gradient energy coefficients chosen to give interfacial energies noted, and stress free transformation strains given in [13, 16, 29].

| Property | Value | Property | Value | Property | Value | Property | Value |
|---|---|---|---|---|---|---|---|
| $C_{11\alpha}$ | 155 GPa | $C_{11\zeta}$ | 168 GPa | $C_{11\gamma}$ | 128 GPa | $C_{11\delta}$ | 63 GPa |
| $C_{22\alpha}$ | 155 GPa | $C_{22\zeta}$ | 168 GPa | $C_{22\gamma}$ | 128 GPa | $C_{22\delta}$ | 63 GPa |
| $C_{12\alpha}$ | 67 GPa | $C_{12\zeta}$ | 89 GPa | $C_{12\gamma}$ | 118 GPa | $C_{12\delta}$ | 28 GPa |
| $C_{33\alpha}$ | 173 GPa | $C_{33\zeta}$ | 195 GPa | $C_{33\gamma}$ | 187 GPa | $C_{33\delta}$ | 65 GPa |
| $C_{13\alpha}$ | 65 GPa | $C_{13\zeta}$ | 67 GPa | $C_{13\gamma}$ | 94 GPa | $C_{13\delta}$ | 44 GPa |
| $C_{23\alpha}$ | $C_{13\alpha}$ | $C_{23\zeta}$ | $C_{13\zeta}$ | $C_{23\gamma}$ | $C_{13\gamma}$ | $C_{23\delta}$ | $C_{13\delta}$ |
| $C_{44\alpha}$ | 40 GPa | $C_{44\zeta}$ | 29 GPa | $C_{44\gamma}$ | 55 GPa | $C_{44\delta}$ | 93 GPa |
| $C_{55\alpha}$ | 40 GPa | $C_{55\zeta}$ | 29 GPa | $C_{55\gamma}$ | 55 GPa | $C_{55\delta}$ | 93 GPa |
| $C_{66\alpha}$ | 44 GPa | $C_{66\zeta}$ | 44 GPa | $C_{66\gamma}$ | 64 GPa | $C_{66\delta}$ | 101 GPa |
| $k_{\alpha\zeta}$ | 0.2 eV/nm | $C_{15\zeta}$ | 23 GPa | $\varepsilon^{00}_{11\gamma}$ | 0.55% | $\varepsilon^{00}_{11\delta}$ | 4.6% |
| $k_{\alpha\gamma}$ | 0.2 eV/nm | $\varepsilon^{00}_{ii\zeta}$ | 2.5% | $\varepsilon^{00}_{22\gamma}$ | 5.64% | $\varepsilon^{00}_{22\delta}$ | 4.6% |
| $k_{\alpha\delta}$ | 0.2 eV/nm | $k_{\zeta\gamma}$ | 0.2 eV/nm | $\varepsilon^{00}_{33\gamma}$ | 5.70% | $\varepsilon^{00}_{33\delta}$ | 7.2% |
| $\sigma_{\alpha\zeta}$ | 0.035 J/m$^2$ | $\sigma_{\zeta\gamma}$ | 0.1 J/m$^2$ | $\sigma_{\gamma\delta}$ | 0.035 J/m$^2$ | $k_{\gamma\delta}$ | 0.2 eV/nm |
| $\sigma_{\alpha\gamma}, \sigma_{\alpha\delta}$ | 0.2 J/m$^2$ | $C_{\zeta}$ | 0.33 | $C_{\gamma}$ | 0.5 | $C_{\delta}$ | 0.6 |
| $A_{\alpha}$ | 100 | $A_{\zeta}$ | 18 | $A_{\gamma}$ | 13 | $A_{\delta}$ | 13 |
| $(B/G)_{\alpha}$ | 2.47 | $(B/G)_{\zeta}$ | 18.34 | $(B/G)_{\gamma}$ | 2.67 | $(B/G)_{\delta}$ | 0.75 |

To investigate the effect of compressive loads during formation of hydrides, simulations were run by applying a displacement boundary condition on the top and bottom boundaries (the radial direction in Fig. 1) to cause a desired compressive strain. Fig. 2b shows reorientation of a single seed after an applied compressive strain of 0.02 in the radial direction. While the initial $\zeta$ hydride transforms to other phases it reorients, and the final table $\delta$ hydride is aligned with the radial direction.

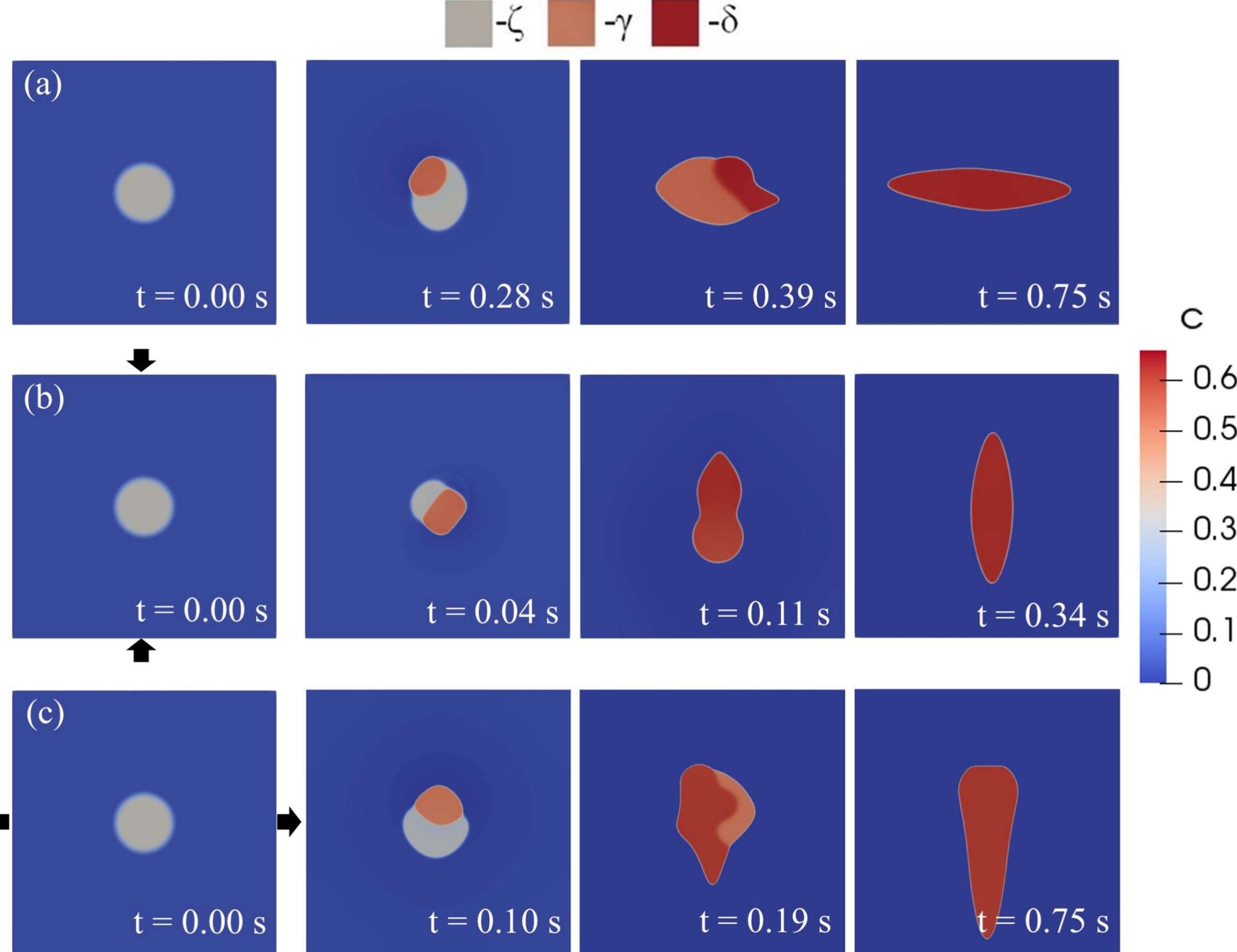

Fig. 2. Single seed of $\zeta$ phase hydride transforms to $\gamma$ then to the stable $\delta$ hydride, (a) no applied strain, (b) and (c) applied strain of 0.02 in the radial and circumferential directions, respectively. The x-axis is the $\left[11\bar{2}0\right]$ direction and the y-axis is the $\left[0001\right]$ direction. Domain size is 75 nm by 75 nm. The colorbar shows the concentration of H.

Hoop stresses occur naturally in cladding tubes due to the internal pressure caused by fission gases created during operation, and additional stresses occur when the claddings are transferred from operation to the storage. Therefore, to determine the strains necessary to cause reorientation of a single hydride, simulations were conducted with applied tensile strains in the circumferential direction ($\left[11\bar{2}0\right]$ direction) by applying a displacement boundary condition on the left and right boundaries of the 2D domain to create the desired strain. Simulation results showed that when the hoop strain of 0.02 is applied, as shown in Fig. 2c, the $\delta$ phase reorients in the radial direction. It should be mentioned that smaller applied strains did not cause reorientation. By comparing the results in Fig. 2b and Fig. 2c, the reorientation process is faster when the compressive radial load is applied causing formation of hydrides parallel to the radial direction, while the tensile hoop strain courses a slower process of reorientation of hydrides in the direction perpendicular to the applied strain.

In the second scenario, we study the case where the $\delta$ hydride has already been formed in the circumferential direction and no reorientation occurred during formation of hydrides, and then an external load is applied (Fig. 3). This simulates the condition where a cladding is in the storage and the hydrides have already formed in the circumferential direction. A single ellipsoidal seed of $\delta$ hydride is placed in middle of the domain. The ellipsoid is oriented in the circumferential direction and it has the aspect ratio of 1:5 to represent the shape of an evolved $\delta$ hydride with no applied load (Fig. 2a). Displacement boundary conditions were appplied similar to the previous cases. Results revealed that the strain required for reorienting the $\delta$ hydride to radial direction is 0.02 for the compressive radial load (Fig. 3) and 0.03 for tensile hoop load (not shown). Fig. 3 also shows the strain maps around the hydride druing the reorientaion, and we will discuss shortly the importance of the strian field aorund the hydrides at the intial stage of loading when the hydride is still oriented in the circumferential direction.

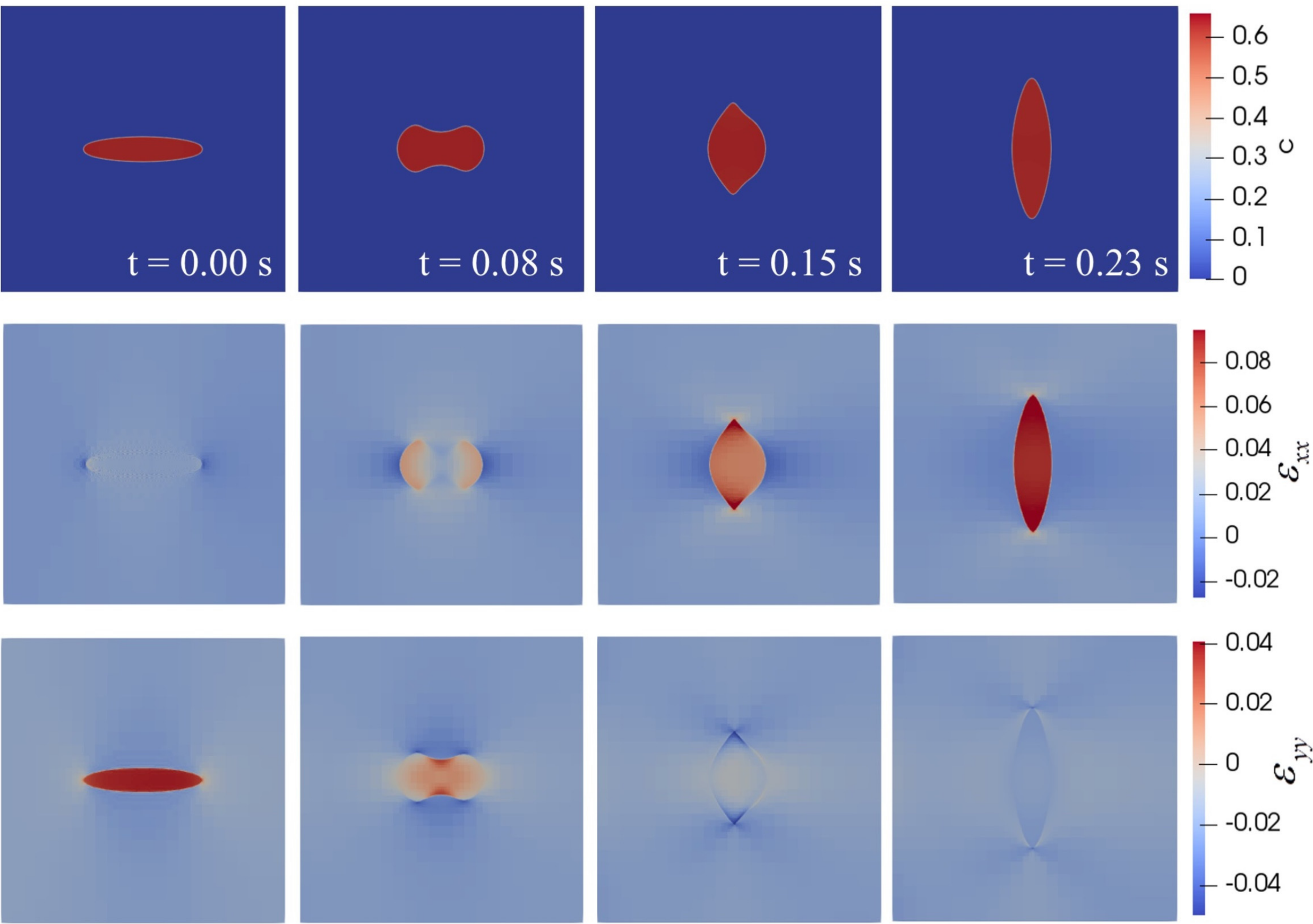

Fig. 3. Single seed of already formed $\delta$ hydride under an applied radial compressive strain (y direction) of 0.02. Domain size is 75 nm by 75 nm. The colorbar shows the concentration of H.

In both of the previous scenarios, the strain necessary to cause reorientation of a single seed in simulations is significantly higher than that in experiments. Colas et al. [3, 4] found the threshold tensile stress necessary to cause reorientation of 40% of hydrides in the cladding to be around 85 MPa, or an applied strain of approximately 0.001 is required. One reason for a higher strain necessary for reorientation in our simulations could be due neglecting the effects of three-dimensional stresses in the 2D simulations. Simulations of $\delta$ phase evolution by Radhakrishnan et al. [30] showed that a 3D single seed of $\delta$ hydride reoriented after an applied strain of 0.01, and in their 2D simulations an applied strain of 0.02 was necessary. However, even a 0.01 strain is one order of magnitude higher than the experimentally determined strain necessary to cause reorientation of hydrides. Another possible reason, which is the focus of this study, is the configuration of hydrides and how they are formed in the vicinity of each other. Other studies have shown that when the hydrides are formed on top of each other, they increase the total energy of the system comparing to the case when they are formed side to side [31]. Furthermore, as the distance between these hydrides decreases, more increase in the total energy of the system is observed.

Our hypothesis is that the configuration of hydrides has a significant influence on the strain field created around hydrides, and this can significantly affect the reorientation process of the neighboring hydrides. However, these effects cannot be seen in single seed simulations. By analyzing the strain map around a hydride, we found that in different stages of hydride formation and growth, tensile strain in the hoop direction and compressive strain in the radial direction exist around the hydride. These strains vary between -0.003 to +0.003 (corresponding to stresses around -400 to +400 MPa) at the distance $a$ from the hydride, where $a$ is the hydride length. In experiments, with only 40% of hydrides reorienting to the radial direction with an applied strain of approximately 0.001, it is entirely possible that some of the hydrides which did not reorient, formed first and created a strain field around them, which in combination with the applied load caused some other hydrides to nucleate and grow in the radial direction. In addition, when two hydrides are formed close enough on top of each other, they will increase the total energy of the system. If we assume that there is an energy threshold at which the reorientation initiates, in the case that the hydrides are formed close to each other, less outside energy, i.e. applied load, would be required to reorient the hydrides. To test this aforementioned hypothesis, several simulations were completed with seeds placed in various distances ($d$) from each other to investigate the effect of neighboring hydrides on the required applied strain to cause reorientation.

Simulations were set up with two seeds placed on top of each other in radial direction with a distance $d$ from each other and under several compressive radial strains. Similar to the one-seed case, two scenarios were considered: transforming from initial $\zeta$ hydride and evolving from already formed $\delta$ hydride. Results of both cases (Fig. 4) revealed that when there are two hydrides in the vicinity of each other, the required load (strain) for reorientation will decrease significantly compared to a single isolated hydride. For instance, for the case of already formed $\delta$ hydrides, the required strain for one single seed of $\delta$ hydride to reorient is 0.02, but when two $\delta$ hydrides are placed close to each other ($d/a$=0.6), the required strain for reorientation drops more than one order of magnitude to 0.0015. To investigate the relation between distance and the required strain, several other simulations were run by varying the distance $d$ and the applied strain required for reorientation was determined for each case. As expected, the results confirmed that as the distance between hydrides decreases, a lesser amount of applied strain/stress is required to initiate the reorientation. This significant decrease in the required strain for reorientation confirms our hypothesis that the configuration of hydrides has a significant effect on the required load for reorientation, and the strain/stress fields created around the hydrides affect the reorientation. It should be mentioned that for some cases where multiphase transformation was present, the seeds did not reorient even though the strains were at the threshold strain for the given seed distance. In these cases, one of the seeds transformed to $\gamma$ before the other one transformed to $\gamma$, absorbing the other seed and leaving just a larger single seed, with no reorient. Nevertheless, the simulations without any phase transformation showed that an applied strain as low as 0.0008 can cause reorientation of formed $\delta$ hydrides.

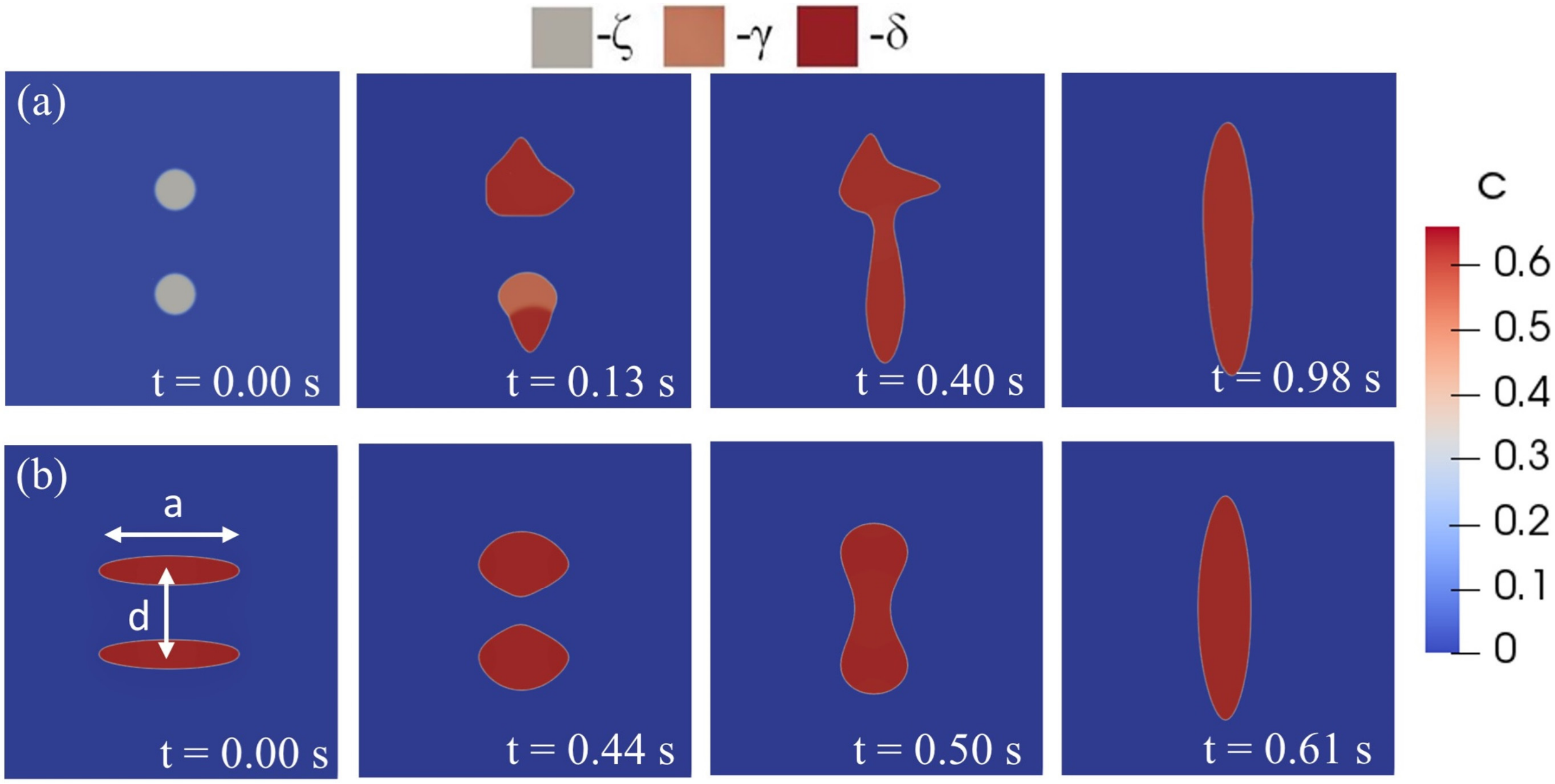

Fig. 4. Two-seed simulations: (a) two $\zeta$ hydride seeds with the distance of $d$=21 nm under compressive strain of 0.002 (radial direction), and (b) two $\delta$ hydrides under compressive strain of 0.0015 (radial direction) and $d/a$=0.6. Domain size is 75 nm by 75 nm. The colorbar shows the concentration of H.

Finally, to represent a case somewhat similar to the real situation, a cluster of hydrides were considered. In this case, five $\delta$ hydrides were placed inside the Zr matrix and then the required strain to cause the reorientation was determined. Simulations for this case (Fig. 5) showed that a strain as low as 0.0007 can make a cluster of hydrides to reorient. This value of strain is in good agreement with the strain value of 0.001 reported in experiments, which again indicates the importance of the configuration of hydrides and its effect on the required strain/stress for reorientation. These necessary applied strains required for reorientation for different cases are summarized plotted in Fig. 6.

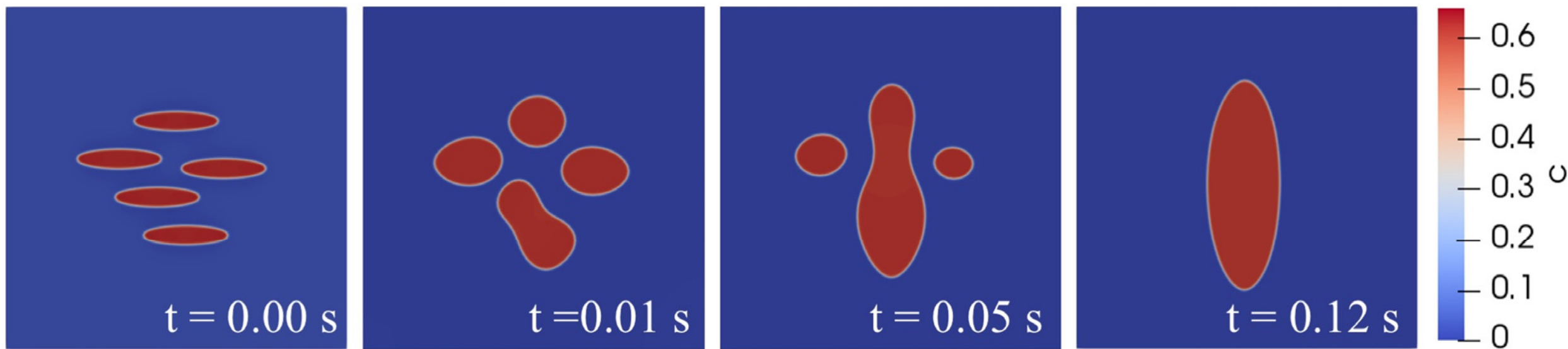

Fig. 5. A cluster of $\delta$ hydrides under applied compressive strain of 0.0007. Domain size is 75 nm by 75 nm. The colorbar shows the concentration of H.

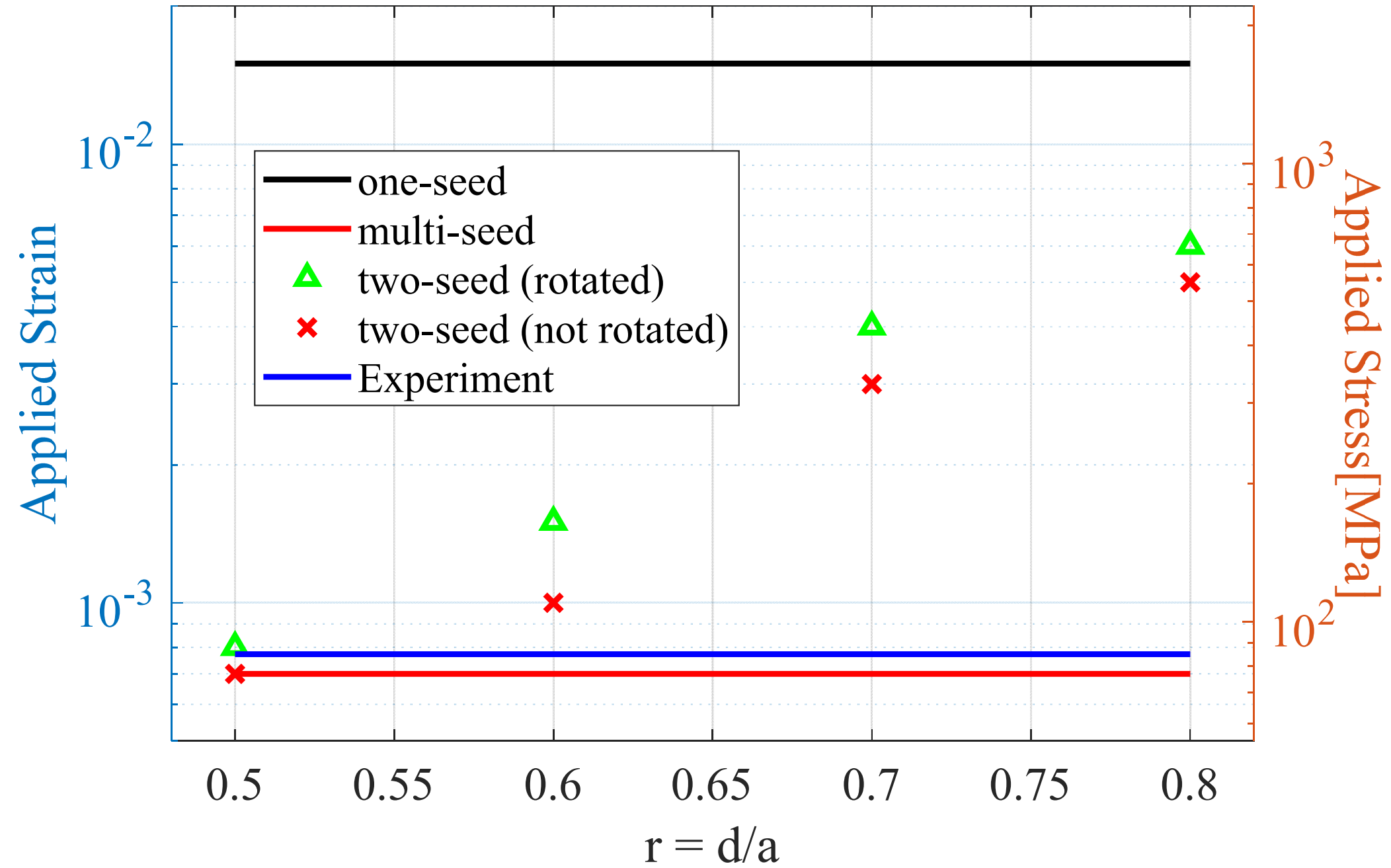

Fig. 6. The minimum applied strain necessary to cause $\delta$ hydride reorientation for different simulation cases compared with that of the experiment [4].

Previous 3D simulations of reorientation of a single hydride needed very high stresses, as high as 9 GPa corresponding to ~8% applied strain, to cause reorientation [15], which is at least one order of magnitude higher than experimental values [4]. From the results in Fig. 6, it can be concluded that strain fields from surrounding hydrides can cause the reorientation of hydrides at applied stress levels much less than 9 GPa and in the order of ~100 MPa. It should be mentioned that the calculated threshold strains versus distance of seeds could be slightly different in 3D simulations, similar to what was seen in single seed simulations [30]. Also, since the applied stain/stress is well below the elastic limit of the material, we expect that the inclusion of dislocations and plastic strain/stress in the model does not meaningfully alter the conclusions of this study, however, to more accurately determine the applied strain/stress necessary for reorientation, simulations should be done in 3D including dislocations and plastic relaxation.

It is also important to notice the significant differences in the time of formation of $\delta$ phase for different cases. When the tensile hoop strain was applied, the seed transformed to the $\delta$ phase much slower than when the radial compressive strain was applied. The radial compressive strain seems to significantly increase the driving force for $\delta$ hydride nucleation from the $\gamma$ hydride phase. It is possible that there is a threshold tensile hoop stress above which $\delta$ hydrides will no longer be the stable phase and $\gamma$ hydrides will become stable. The *B/G* values (Table. 1), which is proportional to ductility of the material, indicate that the $\gamma$ hydrides are more ductile than $\delta$ hydrides. Thus, further investigation of the $\gamma$ hydrides stabilization could be important as they are significantly more ductile than $\delta$ hydrides, and also more ductile than the $\alpha$ Zr, as shown by a density functional theory study [25]. This could potentially help resolving the embrittlement of claddings due to formation of hydrides.

## 4. Conclusion

The effect of applied load on the formation and reorientation of hydrides in nuclear fuel claddings is studied by phase-field modeling study. Single seed simulations showed that to cause significant reorientation of $\delta$ hydrides, an applied strain with a magnitude of at least 0.02 is required. On the other hand, simulations of two-seed hydrides stacked closely on top of each other, and separately simulations of a cluster of hydrides, both showed that the stress fields surrounding the hydrides reduces the necessary strain by more than one order of magnitude; in these cases, a strain as low as 0.0008 can cause the reorientation of hydrides, and the same is also reported in the experimental studies. This confirms the hypothesis that the configuration of hydrides has significant effect on the required load for reorientation, and as hydrides are formed closer to each other, much less strain/stress is required for reorientation. Distance between hydrides was shown to be a significant factor in determining the threshold strain necessary for reorientation, and as this distance decreases the required strain decrease accordingly. In addition, when several hydrides are in the proximity of each other, the process of reorientation of several hydrides includes the growth of smaller precipitates on top of one another forming a larger hydride aligned in the radial direction. The coalescence of smaller hydrides forming larger hydrides assists the reorientation process, lowering the required applied load for reorientation.

Simulations also showed that the tensile hoop stresses significantly reduce the driving force for the transformation of $\gamma$ to $\delta$ hydrides. This may indicate that $\gamma$ hydrides can be made stable under some specific applied stress conditions. The possibility of stabilizing the $\gamma$ phase should be studied further both experimentally and computationally to determine if it could reduce the risk of hydride embrittlement in cladding materials since $\gamma$ hydrides are much more ductile than $\delta$ hydrides.

**Acknowledgment**

We would like to thank the MOOSE team in Idaho National Laboratory for their support in debugging the code.

## CRediT Author Statement

**Alireza Toghraee:** Conceptualization, Methodology, Software, Formal analysis, Writing- Original draft preparation. **Jacob Bair:** Conceptualization, Methodology, Software, Formal analysis, Writing- Original draft preparation. **Mohsen Asle Zaeem:** Supervision, Conceptualization, Methodology, Formal analysis, Writing- Reviewing and Editing, Funding Acquisition.

## Declaration of Competing Interest

The authors declare that they have no known competing financial interests or personal relationships that could have appeared to influence the work reported in this paper.

## Date Availability

Data and results will be made available upon reasonable requests.